\documentclass[
 reprint,
 amsmath,amssymb,
 aps,
 prl,
]{revtex4-2}

\usepackage{graphicx}
\usepackage{dcolumn}
\usepackage{bm}
\usepackage{hyperref}
\usepackage[utf8]{inputenc}
\usepackage{enumitem}
\usepackage{cancel}

\usepackage{color}

\newcommand{\ket}[1]{\left| #1 \right\rangle}

\newcommand{\vev}[1]{\langle #1 \rangle}

\begin{document}

\title{Neutrino Oscillations at JUNO, the Born Rule,\\and Sorkin's Triple Path Interference}

\author{Patrick Huber}
\email{pahuber@vt.edu}

\author{Hisakazu Minakata}
\email{hisakazu.minakata@gmail.com}

\author{Djordje Minic}
\email{dminic@vt.edu}

\author{Rebekah Pestes}
\email{rebhawk8@vt.edu}

\author{Tatsu Takeuchi}
\email{takeuchi@vt.edu}

\affiliation{Center for Neutrino Physics, Department of Physics, Virginia Tech, Blacksburg VA 24061, USA}

\begin{abstract}
We argue that neutrino oscillations at JUNO offer a unique opportunity to study
Sorkin's triple-path interference, which is predicted to be zero
in canonical quantum mechanics by virtue of the Born rule. 
In particular, we compute the expected bounds on triple-path interference at JUNO 
and demonstrate that they are comparable to those already available from electromagnetic probes.
Furthermore, the neutrino probe of the Born rule is much more direct due to an intrinsic independence from any boundary conditions,
whereas such dependence on boundary conditions is always present in the case of electromagnetic probes.
Thus, neutrino oscillations present an ideal probe of this aspect of the foundations of quantum mechanics.
\end{abstract}


\maketitle

\section{Introduction}

Obtaining a deeper understanding of Quantum Mechanics (QM) is homework leftover from the 20th century.
The question is becoming more acute with the development of QM based technologies
already impacting our everyday lives (semi-conductors, super-conductors, etc.)
as well as the promise of various quantum information technologies that may be realized in the
not too distant future \cite{PRXQuantum.1.020101}.
In this letter, we emphasize the relevance of neutrino physics to address 
various foundational questions in QM. 
In particular, we consider the potential of neutrino oscillations to
probe the triple-path interference of Sorkin \cite{Sorkin:1994dt} as a direct test of the 
Born rule and compare the expected bound to those currently available from electromagnetic experiments.

There are many features that distinguish QM from classical mechanics (CM).
Though the statistical nature of QM as opposed to the deterministic nature of CM
is often emphasized in textbooks, many other differences exist as well; 
for instance, in correlations \cite{Bell:1964kc,Bell:1987hh,Clauser:1969ny},
and in the presence or absence of interference.
However, it has been noted that not only do QM correlations go beyond those of CM but 
are themselves restricted \cite{Cirelson:1980ry,Landau:1987},
and are not as large as that allowed by logic and relativity \cite{Popescu:1994}.
QM interference is also restricted in that the Born rule only allows for pair-wise interference
between paths, but not for triple-path interference or higher \cite{Sorkin:1994dt}.

The fact that canonical QM itself is restricted, with no apparent physical or logical reason,
points to the possible existence of consistent theories that go beyond QM boundaries, perhaps at the expense
of some of our cherished principles which we currently hold to be fundamental.
It is up to experiments to determine whether Nature always stays within those QM boundaries,
or occasionally ventures outside, and under what conditions.
Indeed, it has been argued that it would (or should) in the context of quantum gravity and cosmology \cite{Penrose:2014nha,Gell-Mann:2013hza}, 
or in the realm of quantum measurement \cite{Weinberg:2016axv}, 
or in the domain of macroscopic quantum systems \cite{Leggett:2007zz}. 

The celebrated Bell's inequality \cite{Bell:1964kc,Bell:1987hh} (1964) and its
generalization by Clauser, Horne, Shimony, and Holt, the CHSH inequality \cite{Clauser:1969ny} (1969),
provide bounds that CM correlations cannot violate.
That experimental correlations violate these bounds has been confirmed by
various groups using photons \cite{Aspect:1981zz} (1981),
ions and atoms, Josephson junctions, and NV centers in diamond (See FIG.~1 of \cite{PhysRevLett.115.180408}).
However, Tsirelson has shown that the CHSH correlator is bounded 
from above for QM as well \cite{Cirelson:1980ry,Landau:1987} (1980) and while this bound is naturally larger than that of CM, it does not saturate the logically allowed maximum,
despite the fact that such a saturation does not contradict relativistic causality as demonstrated
by Popescu and Rohrlich \cite{Popescu:1994} (1994).
Whether experimental correlations violate the QM Tsirelson bound or not 
is being checked using photons \cite{PhysRevLett.115.180408,Tian:2020}.

The Bell/CHSH inequalities involve correlations between entangled pairs of spatially separated observables. 
In contrast, the Leggett-Garg inequalities (LGI's) \cite{Leggett:1985zz} (1985) involve
correlations between temporary separated measurements of a single observable and are expected to hold
for ``macroscopic'' systems.
Since QM satisfies neither of the two assumptions underlying the LGI's, 
namely {macroscopic realism} and {non-invasive measurability},
QM correlations can be expected to violate them, though there is some subtlety 
in how those correlations should be defined in QM to reflect the setup of each experiment.
Nevertheless, experimental checks that demonstrate the violation of some form of the LGI
have been performed using a variety of systems including superconducting qubits, nuclear spins,
photons (see Table 1 of Ref.~\cite{Emary_2013,Emary_2014}),
and more recently using the neutrino oscillation data from Minos \cite{Formaggio:2016cuh} (2016) and Daya-Bay \cite{Fu:2017hky} (2017).

Neutrino experiments are, in many respects, ideal laboratories for foundational quantum research
given the long coherence times that neutrino states have, and the fundamental role of interference in 
neutrino oscillation phenomena.
In addition to probing the LGI's \cite{Formaggio:2016cuh,Fu:2017hky}, 
the violation of which is not surprising, 
the potential of neutrino oscillations to constrain models that generalize and go beyond canonical QM 
have also been explored.
For instance, Refs.~\cite{Donadi:2013uxa,Bahrami:2013hta} investigate whether
neutrino oscillations can constrain the Continuous Spontaneous Localization model \cite{Ghirardi:1989cn} (expected effect is too small to be observed),
while Ref.~\cite{Minic:2020zjb} argues that atmospheric neutrino data can constrain
Nambu QM extensions \cite{Minic:2002pd}.
The possibility of utilizing M\"ossbauer neutrinos to probe the
time-energy uncertainty relation has been discussed in Ref.~\cite{Raghavan:2012sy}.

In this letter, we explore the potential of the neutrino oscillation experiment JUNO \cite{An:2015jdp} 
to look for the triple-path interference of Sorkin \cite{Sorkin:1994dt} (1994).
Sorkin provides a classification scheme for theories that go beyond CM 
in terms of the existence of multi-path interferences.
Only double-path interferences exist in QM due to the Born rule
though, in principle, triple and higher-order interferences are possible. 
Experimental bounds have been placed on the presence of triple-path interference 
using photons \cite{Sinha:2010} (2010) and Liquid State NMR \cite{Park:2012} (2012).
Both experiments report upper bounds on the ratio of triple-path to double-path interferences 
of order $10^{-2}$.
Improving the photon bound requires controlling the sensitivity of multi-slit interference to
the change in the boundary condition due to the opening and closing of slits \cite{Yabuki:1986,PhysRevA.85.012101,PhysRevLett.113.120406,Sinha:2015,Rengaraj:2018}.
Neutrino oscillation, on the other hand, does not involve any slits and
always has three mass eigenstates interfering with each other.
However, at different baselines, neutrino energies, and neutrino energy resolutions, 
it effectively becomes a double-path interference experiment due to 
the large separation in scale between $\delta m^2_{21}$ and $\delta m^2_{31}$.
Indeed, JUNO \cite{An:2015jdp} is expected to be the first experiment in which the interference between
the atmospheric and solar oscillation amplitudes is clearly visible \cite{Huber:2019frh}.

In the following, we first review Sorkin's definition of multi-path interference and
the classification of theories based on their presence or absence \cite{Sorkin:1994dt}.
We then review our previous work from Ref.~\cite{Huber:2019frh},
which looked at the potential of JUNO to detect deviations of the
neutrino oscillation probabilities from their QM predictions.
This is effectively the same problem as we are considering in this letter,
albeit imposing a particular normalization for the possible triple-path interference.
The expected bounds on triple-path interference at JUNO for several other normalization/parametrization
choices with details of the analyses are presented next.
The parametrizations are chosen under the caveat that the contribution
of the triple-path interference should be invisible to pre-JUNO experiments,
and includes the one which facilitates comparison of the bound with the photon/NMR results.
We conclude with a discussion on 
the necessity of a concrete model which predicts triple-path interference to 
derive a more physically meaningful bound.

\noindent

\section{The Born Rule and Sorkin's Multi-Path Interference}

We follow the discussion of Sorkin in Ref.~\cite{Sorkin:1994dt}.
Let 
\begin{equation}
P_{n}(A,B,C,\cdots)
\end{equation}
denote the probability of a system to go from an initial state $\ket{\alpha}$ to a final state $\ket{\beta}$
when $n$ pathways $A,B,C,\dots$ connecting the two are available.
Classically, we have
\begin{equation}
P_{n}(A,B,C,\cdots) \,=\, P_{1}(A) + P_{1}(B) + P_{1}(C) + \cdots\;,
\end{equation}
for any number of paths.
Quantum mechanically, we have for two paths
\begin{eqnarray}
P_{2}(A,B) 
& = & |\psi_A + \psi_B|^2 \vphantom{\Big|}\cr 
& = &
\underbrace{|\psi_A|^2}_{\displaystyle P_{1}(A)} + 
\underbrace{|\psi_B|^2}_{\displaystyle P_{1}(B)} + 
\underbrace{(\psi_A^*\psi_B^{\phantom{*}} + \psi_B^*\psi_A^{\phantom{*}})}_{\displaystyle I_{2}(A,B)}
\;.
\end{eqnarray}
The extra term
\begin{equation}
I_{2}(A,B) \;=\; P_{2}(A,B)-P_{1}(A)-P_{1}(B) 
\end{equation}
is the ``interference'' of the two paths $A$ and $B$.
The non-vanishing of this double-path interference, 
$I_{2}(A,B)\neq 0$, distinguishes QM from CM.

In QM the superposition principle allow us to superimpose
an arbitrary number of ``paths'' on top of each other. 
Indeed, in the path integral approach we superimpose an infinite number of them \cite{Feynman_Hibbs}.
However, the Born Rule dictates that all the superimposed paths only interfere with each other 
in a pairwise manner.
For instance, for three paths we have
\begin{eqnarray}
\lefteqn{
P_{3}(A,B,C) 
\;=\; |\psi_A + \psi_B + \psi_C|^2 
\vphantom{\Big|}
}\cr
& = & 
 \underbrace{|\psi_A|^2}_{\displaystyle P_{1}(A)} 
+\underbrace{|\psi_B|^2}_{\displaystyle P_{1}(B)} 
+\underbrace{|\psi_C|^2}_{\displaystyle P_{1}(C)}
+\underbrace{(\psi_A^*\psi_B^{\phantom{*}} + \psi_B^*\psi_A^{\phantom{*}})}_{\displaystyle I_{2}(A,B)}
\cr
& &
+\underbrace{(\psi_B^*\psi_C^{\phantom{*}} + \psi_C^*\psi_B^{\phantom{*}})}_{\displaystyle I_{2}(B,C)}
+\underbrace{(\psi_C^*\psi_A^{\phantom{*}} + \psi_A^*\psi_C^{\phantom{*}})}_{\displaystyle I_{2}(C,A)}
\vphantom{\Big|}
\cr
& = & P_{2}(A,B) + P_{2}(B,C) + P_{2}(C,A) \vphantom{\Big|}\cr
& &
- P_{1}(A) - P_{1}(B) - P_{1}(C)
\;.\vphantom{\Big|}
\end{eqnarray}
Only pairwise interferences between the pairs $(A,B)$, $(B,C)$, and $(C,A)$ appear.
Therefore, it makes sense to define any deviation from this relation as the
triple-path interference:
\begin{eqnarray}
\lefteqn{I_{3}(A,B,C) \vphantom{\Big|}}
\cr
& = & P_{3}(A,B,C)
-P_{2}(A,B)
-P_{2}(B,C)
-P_{2}(C,A)
\cr
& &
+\;P_{1}(A)
+P_{1}(B)
+P_{1}(C)
\;.
\vphantom{\Big|}
\end{eqnarray}
For both CM and QM, this triple-path interference is zero for any triplet of
paths.

In a similar fashion, the $n$-path interference for $n\ge 4$ can be defined as
\begin{eqnarray}
\lefteqn{I_{n}(A_1,A_2,\cdots,A_n) \vphantom{\Big|}} \cr
& = & P_{n}(A_1,A_2,\cdots,A_n) 
-\sum P_{n-1}(A_i,A_j,\cdots) 
\cr
& & +\sum P_{n-2}(A_i,\cdots) - \cdots - (-1)^n \sum P_1(A_i)
\;,\qquad
\end{eqnarray}
which are always zero for both CM and QM. 
Therefore, CM and QM can be characterized by 
\begin{eqnarray}
\mathrm{CM} & : & \mbox{$I_n = 0$ for $n\ge 2$} \;,\vphantom{\Big|}\cr
\mathrm{QM} & : & \mbox{$I_2\neq 0$, $I_n = 0$ for $n\ge 3$} \;.\vphantom{\Big|}
\end{eqnarray}
Experimental confirmation of $I_3=0$ would be a confirmation of the Born rule.
In Refs.~\cite{Sinha:2010,Park:2012}, bounds were placed on the parameter
\begin{equation}
\kappa \;=\; \dfrac{\varepsilon}{\delta}\;,
\end{equation}
where 
\begin{eqnarray}
\varepsilon & = & I_3(A,B,C) \;,\vphantom{\Big|}\cr
\delta & = & |I_2(A,B)| + |I_2(B,C)| + |I_2(C,A)| \;.\vphantom{\Big|}
\label{epsilondeltadef}
\end{eqnarray}
Ref.~\cite{Sinha:2010} reports $\kappa=0.0064\pm 0.0119$ for a multi-slit experiment
with a single photon source, while Ref.~\cite{Park:2012} reports $\kappa=0.007\pm 0.003$
based on a liquid state NMR experiment.
Thus, the 1$\sigma$ deviation of $\kappa$ from zero allowed by these experiments
is $|\kappa| < 0.01\sim 0.02$.

\section{Huber, Minakata, and Pestes}

Here, we review the analysis of Ref.~\cite{Huber:2019frh}.
According to canonical QM, the neutrino oscillation amplitude for $\nu_\beta\to\nu_\alpha$ 
at distance $x$ from the source is given by
the superposition of the contributions of the three mass eigenstates:
\begin{eqnarray}
\lefteqn{S_{\alpha\beta}^{(123)} 
\;=\; 
 U_{\alpha 1}U^*_{\beta 1}
+U_{\alpha 2}U^*_{\beta 2}e^{-i\Delta_{21}x}
+U_{\alpha 3}U^*_{\beta 3}e^{-i\Delta_{31}x} 
}
\vphantom{\Big|}\cr 
& = & 
\underbrace{
\left(U_{\alpha 1}U^*_{\beta 1}+U_{\alpha 2}U^*_{\beta 2}+U_{\alpha 3}U^*_{\beta 3}\right)
}_{\displaystyle \delta_{\alpha\beta}}
\vphantom{\Big|}\cr
& & 
+ \underbrace{U_{\alpha 2}U^*_{\beta 2}\left(e^{-i\Delta_{21}x}-1\right)}_{\displaystyle S_{\alpha\beta}^{\mathrm{sol}}}
+ \underbrace{U_{\alpha 3}U^*_{\beta 3}\left(e^{-i\Delta_{31}x}-1\right)}_{\displaystyle S_{\alpha\beta}^{\mathrm{atm}}}
\cr
& = & \delta_{\alpha\beta} + S_{\alpha\beta}^{\mathrm{sol}} + S_{\alpha\beta}^{\mathrm{atm}}
\;,\vphantom{\Big|}
\end{eqnarray}
where
\begin{equation}
\Delta_{ij} \;=\; \dfrac{\delta m_{ij}^2}{2E}\;.
\end{equation}
The Born rule gives the QM oscillation probability for $\nu_\beta\to\nu_\alpha$ as
\begin{eqnarray}
\lefteqn{P_{\text{QM}}(\nu_\beta\to\nu_\alpha)
\;=\; \big|S_{\alpha\beta}^{(123)}\big|^2
}
\vphantom{\bigg|}\cr
& = & \underbrace{
\delta_{\alpha\beta} + \bigl|S_{\alpha\beta}^{\mathrm{sol}}\bigr|^2 + \bigl|S_{\alpha\beta}^{\mathrm{atm}}\bigr|^2
+ 2\delta_{\alpha\beta}\,\Re\!
\left(S_{\alpha\beta}^{\mathrm{sol}} + S_{\alpha\beta}^{\mathrm{atm}} 
\right)}_{\displaystyle P_{\beta\alpha}^{\text{non-int-fer}}}
\cr
& & + \underbrace{
\left(S_{\alpha\beta}^{\mathrm{sol}*}S_{\alpha\beta}^{\mathrm{atm}}+S_{\alpha\beta}^{\mathrm{atm}*}S_{\alpha\beta}^{\mathrm{atm}}\right)}_{\displaystyle P_{\beta\alpha}^{\text{int-fer}}}
\cr
& = & P_{\beta\alpha}^{\text{non-int-fer}} + P_{\beta\alpha}^{\text{int-fer}} \;.
\end{eqnarray}
This is the QM prediction.
To check this relation, Ref.~\cite{Huber:2019frh} introduces the parameter $q$ as
\begin{equation}
P_{\text{exp}}(\nu_\beta\to\nu_\alpha)
\;=\; P_{\beta\alpha}^{\text{non-int-fer}} + q P_{\beta\alpha}^{\text{int-fer}}
\;,
\label{qdef}
\end{equation}
and discusses the bounds that can be placed on $q$ by the JUNO experiment \cite{An:2015jdp},
that is,
the expected $P_{\text{exp}}(\nu_\beta\to\nu_\alpha)$ at JUNO assuming canonical QM is simulated using 
GLoBES \cite{Huber:2004ka,Huber:2007ji} to which the right-hand side of Eq.~\eqref{qdef}
is fit to calculate the expected bound on $q$.
Since the $P_{\beta\alpha}^{\text{non-int-fer}}$ term
includes interference effects between the $1$-$2$ and $1$-$3$ mass eigenstates, 
the parameter $q$ checks for $2$-$3$ interference.
However, any deviation of $P_{\text{exp}}(\nu_\beta\to\nu_\alpha)$ from the QM prediction can also be
interpreted as due to Sorkin's triple-path interference.
Indeed, Eq.~\eqref{qdef} effectively parametrizes the size of triple-path interference as
\begin{eqnarray}
\varepsilon & = &
P_{\text{exp}}(\nu_\beta\to\nu_\alpha) - P_{\text{QM}}(\nu_\beta\to\nu_\alpha)
\vphantom{\Big|}\cr
& = & (q-1) P_{\beta\alpha}^{\text{int-fer}}\;.
\label{HMPepsilon}
\end{eqnarray}
Note that this parametrization of the triple-path interference renders it
invisible to all pre-JUNO experiments.
Thus the bound on $q-1$ can be reinterpreted as a bound on $\varepsilon$
normalized to $P_{\beta\alpha}^{\text{int-fer}}$.  
Using Wilks' Theorem \cite{Wilks:1938}, 
we find that the analysis of Ref.~\cite{Huber:2019frh} imposes
a $1\sigma$ allowed range on $q-1$ given by
\begin{equation}
-0.17 < (q-1) <  0.12
\end{equation}
Note that in the absence of an actual theory with triple-path interference 
which predicts the energy and baseline dependence of $\varepsilon$, 
we must make a somewhat arbitrary choice as in Eq.~\eqref{HMPepsilon}.
By fitting the data with constant $\kappa = \varepsilon/\delta$,
Refs.~\cite{Sinha:2010,Park:2012} are assuming that
$\varepsilon \propto \delta$, cf. Eq.~\eqref{epsilondeltadef}, which is another arbitrary choice.
However, $\delta$ does have the advantage over $P_{\beta\alpha}^{\text{int-fer}}$
in that all three paths are treated equally.
Due to this, and also for the ease of comparison with the photon and NMR results,
we redo the analysis of \cite{Huber:2019frh} with this normalization.

\section{Possible JUNO bound on $\kappa$}

The parameter $\kappa$ is introduced as
\begin{eqnarray}
\varepsilon
& = & P_{\text{exp}}(\nu_\beta\to\nu_\alpha) - P_{\text{QM}}(\nu_\beta\to\nu_\alpha)
\vphantom{\Big|}\cr
& = & \kappa\,\Bigl( |I_{\alpha\beta}(1,2)| + |I_{\alpha\beta}(1,3)| + |I_{\alpha\beta}(2,3)| \Bigr)
\;,
\vphantom{\Big|}
\label{eq:kappainterference}
\end{eqnarray}
where
\begin{equation}
I_{\alpha\beta}(i,j) \;=\; 2\,\Re
\left(U_{\alpha i}^{\phantom{*}}U_{\beta i}^* U_{\alpha j}^* U_{\beta j}^{\phantom{*}}\;e^{-i\Delta_{ij}x}\right)
\end{equation}
is the interference between the $i$-th and $j$-th mass eigenstates.
Strictly speaking, one should perform a global fit to all neutrino oscillation experiments
to bound $\kappa$.  However, $|I_{\alpha\beta}(2,3)|$ was invisible to all pre-JUNO 
experiments and would have averaged to a small constant, while the other terms
would have been absorbed into the uncertainties in the mixing angles.
This justifies our JUNO-only analysis.

We use GLoBES \cite{Huber:2004ka,Huber:2007ji} to simulate the JUNO experiment as detailed in Ref.~\cite{Huber:2019frh}.
The simulation is set up with two detectors: a JUNO-like far detector with a
fiducial mass of $20\,\text{kt}$ and an energy resolution of
$3\%/\sqrt{E}$ at a distance of $53\,\text{km}$ from a nuclear reactor
source with a total power of $36\,\text{GWth}$, and a TAO-like
\cite{Sisti:2020jie} near detector with a fiducial mass of
$1\,\text{ton}$ and an energy resolution of $1.7\%/\sqrt{E}$ at a
distance of $30\,\text{m}$ from a $4.6\,\text{GWth}$ nuclear reactor
core; we assume a total data taking time of 6 years. 

\begin{figure}
	\centering
	\includegraphics[width=8cm]{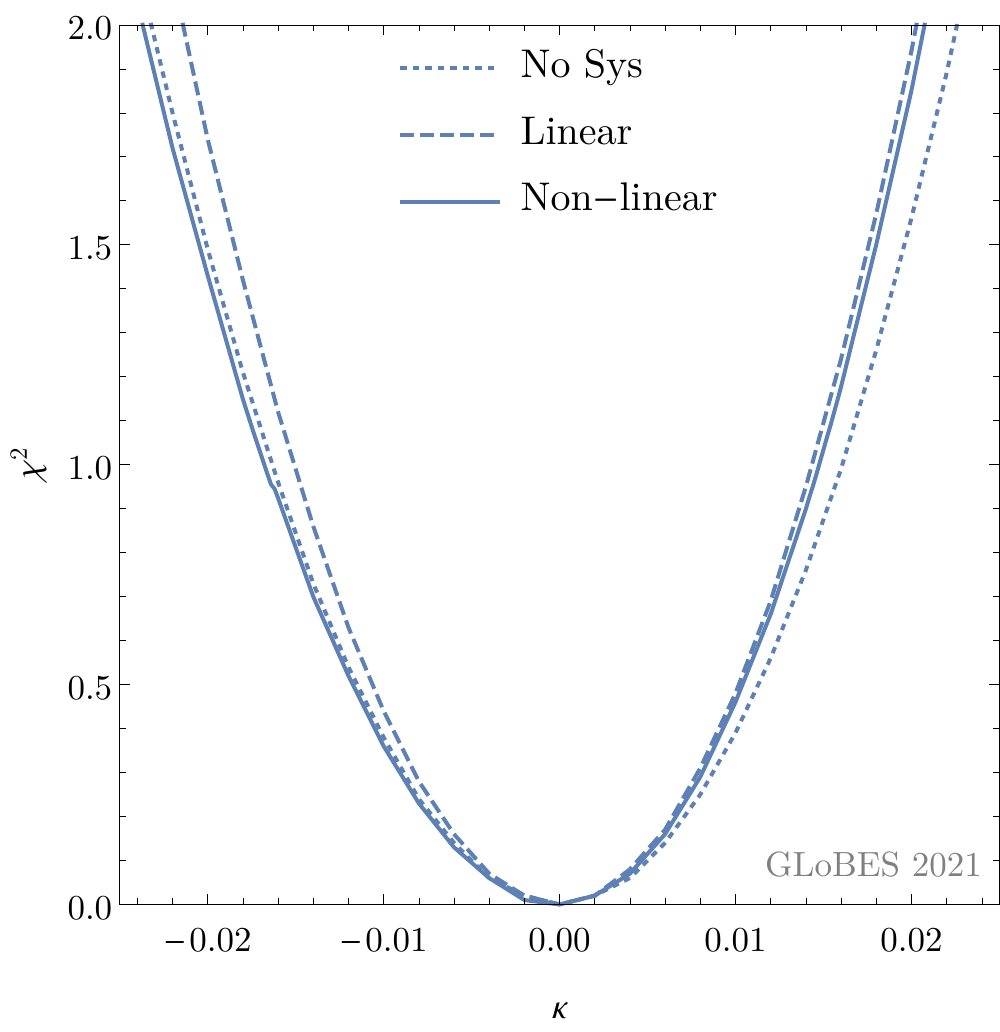}
	\caption{$\chi^2$ curves for $\kappa$ as defined in Eq.~\ref{eq:kappainterference}.  
The (a) solid, (b) dashed, and (c) dotted curves correspond to simulations in which the uncertainties are treated 
differently as detailed in the main text.}
	\label{fig:kappachi2}
\end{figure}

For the purpose of producing the simulated data $P_\text{exp}(\nu_\beta\to\nu_\alpha)$, we assume 
canonical QM with normal ordering 
to be the true mass ordering and the relevant oscillation parameters to be
$\Delta m^2_{21}=7.54\times 10^{-5}\,\text{eV}^2$, 
$\Delta m^2_{31}=2.43\times 10^{-3}\,\text{eV}^2$, 
$\theta_{12}=33.6^{\circ}$, and
$\theta_{13}=8.9^{\circ}$
\cite{deSalas:2020pgw}.
The theoretical QM rates $P_\text{QM}(\nu_\beta\to\nu_\alpha)$ are calculated with the same
inputs and the difference between $P_\text{exp}(\nu_\beta\to\nu_\alpha)$ and $P_\text{QM}(\nu_\beta\to\nu_\alpha)$
is fit with Eq.~\eqref{eq:kappainterference}.
The results of our simulation are shown in Fig.~\ref{fig:kappachi2} for the following three analyses:

\renewcommand{\labelenumi}{(\theenumi)}
\renewcommand{\theenumi}{\alph{enumi}}
\begin{enumerate}[leftmargin=*]
\item (solid line)
For each detector, we use a model for non-linear effects in the reconstruction
of the positron energy as described in Ref.~\cite{Forero:2017vrg}, which includes terms up to the cubic in the positron energy.
To account for the uncertainties in the reactor anti-neutrino flux prediction, we
conservatively introduce a nuisance parameter to each of our 100
energy bins with the spectrum computed before applying the energy resolution function. 
This is equivalent to the assumption of \emph{no}
prior knowledge of fluxes, as in Ref.~\cite{Forero:2017vrg}. 

\item (dashed line)
The same analysis repeated except assuming that the
energy calibration error for each detector is linear.

\item (dotted line) 
Simulation without a near detector while assuming perfect knowledge of
detector and source systematics.

\end{enumerate}

An interesting feature of Fig.~\ref{fig:kappachi2} is that analysis (c) with no systematic uncertainties provides a weaker constraint for $\kappa$ than analyses (a) and (b) with systematic uncertainties and a near detector. 
This demonstrates that the presence of a near detector not only constrains the neutrino flux, but
also provides extra constraints on possible deviations of the neutrino oscillation probabilities from canonical QM.
Using Wilks' Theorem, the $1\sigma$ allowed range of $\kappa$ for analysis (a) is found to be
\begin{equation}
-0.017 < \kappa < 0.015
\end{equation}
%

\section{Constraining Other Forms of the Interference Term}

\begin{figure}
	\centering
	\includegraphics[width=8cm]{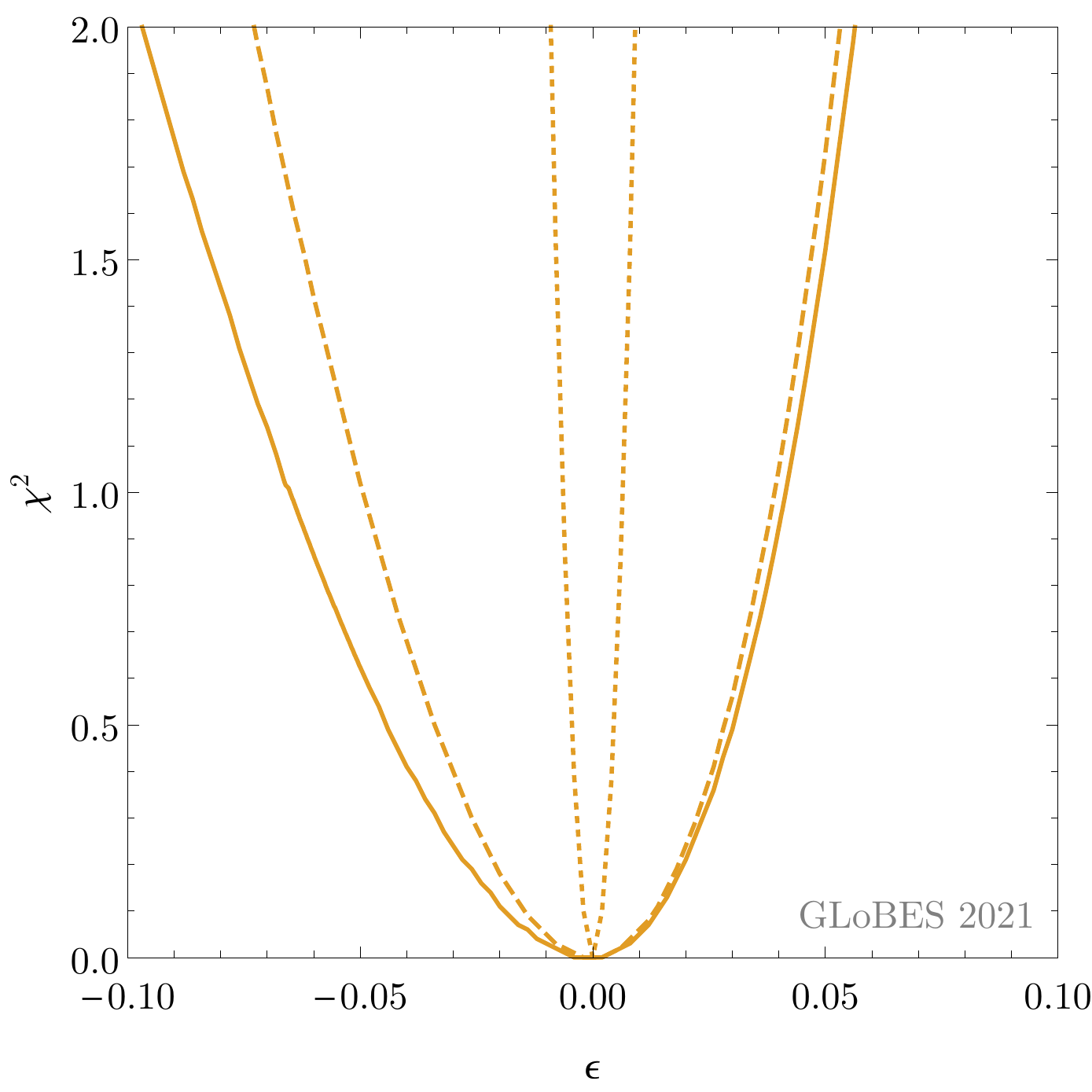}
	\caption{$\chi^2$ curves for $\varepsilon$ as a constant.  
	The (a) solid, (b) dashed, and (c) dotted curves correspond to simulations in which the uncertainties are treated 
differently as detailed in the main text.}
	\label{fig:epsilonchi2}
\end{figure}

\begin{figure}
	\centering
	\includegraphics[width=8cm]{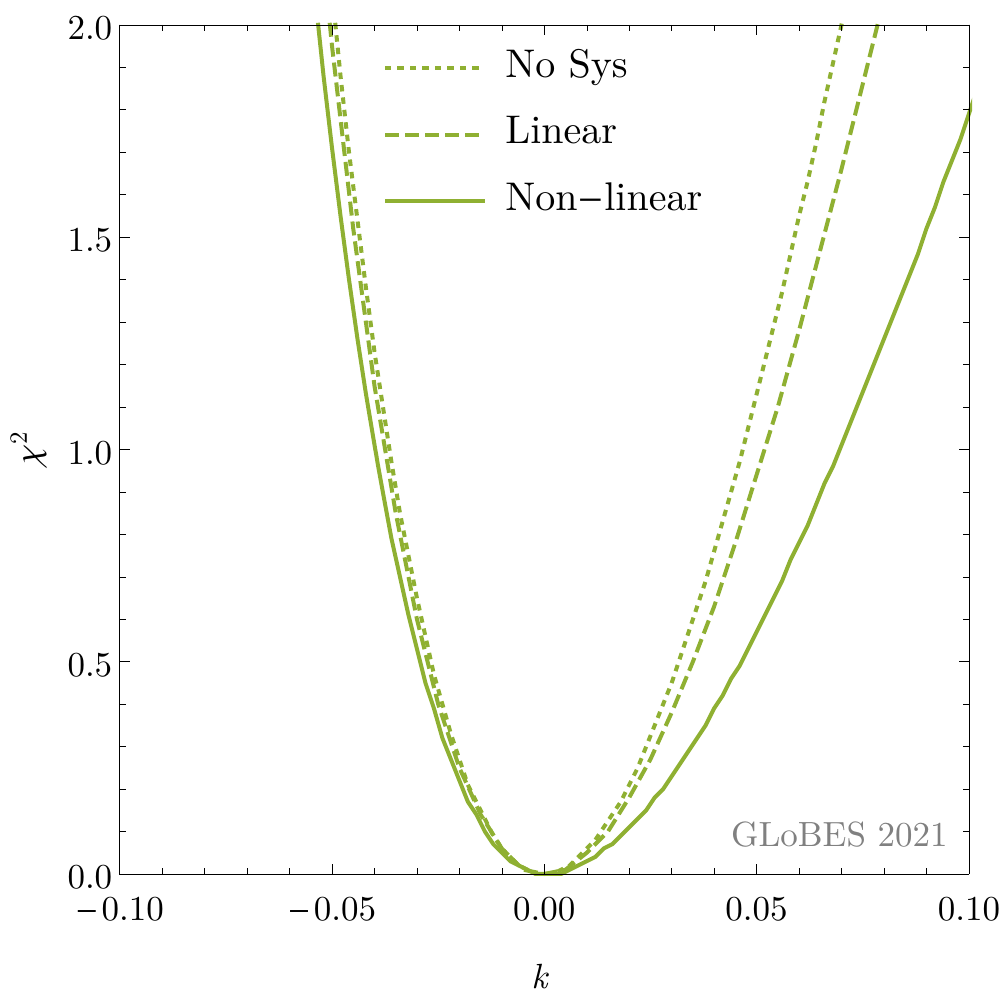}
	\caption{$\chi^2$ curves for $k$ as defined in Eq.~\ref{eq:kinterference}.  
	The (a) solid, (b) dashed, and (c) dotted curves correspond to simulations in which the uncertainties are treated 
differently as detailed in the main text.}
	\label{fig:kchi2}
\end{figure}

In addition to Eq.~\eqref{eq:kappainterference}, which we will refer to as case (1),
we consider two other forms for the triple-path interference $\varepsilon$:
(2) constant $\varepsilon$, \textit{i.e.} $\varepsilon$ independent of $L/E$, 
and (3) $\varepsilon$ proportional $1-P_\text{QM}(\nu_\beta\to\nu_\alpha)$,
that is
\begin{eqnarray}
\varepsilon & = & 
P_{\text{exp}}(\nu_\beta\to\nu_\alpha) - P_{\text{QM}}(\nu_\beta\to\nu_\alpha)
\vphantom{\Big|}\cr
& = &
k\,(1-P_\text{QM}(\nu_\beta\to\nu_\alpha))\,,
	\label{eq:kinterference}
\end{eqnarray}
where $k$ is a constant.
In case (2), we are considering the possibility that triple-interference
is hidden in the uncertainty of the overall count rate, while 
in case (3) we are assuming
\begin{equation}
P_{\text{exp}}(\nu_\beta\to\nu_\alpha)
\;=\; k + (1-k)P_{\text{QM}}(\nu_\beta\to\nu_\alpha)
\;.
\end{equation}
For each case, we perform the same three analyses as in the previous section and plot the results in Figs.~\ref{fig:epsilonchi2} and \ref{fig:kchi2}, respectively.  
For case (2), the $1\sigma$ allowed range for $\varepsilon$ from analysis (a) is
\begin{eqnarray}
-0.065 \;<\; & \varepsilon & \;<\; 0.042 
\end{eqnarray}
while for case (3), the bound on $k$ from analysis (a) is
\begin{eqnarray}
-0.040 \;<\; & k & \;<\; 0.069
\end{eqnarray}

\begin{figure}
	\centering
	\includegraphics[width=8cm]{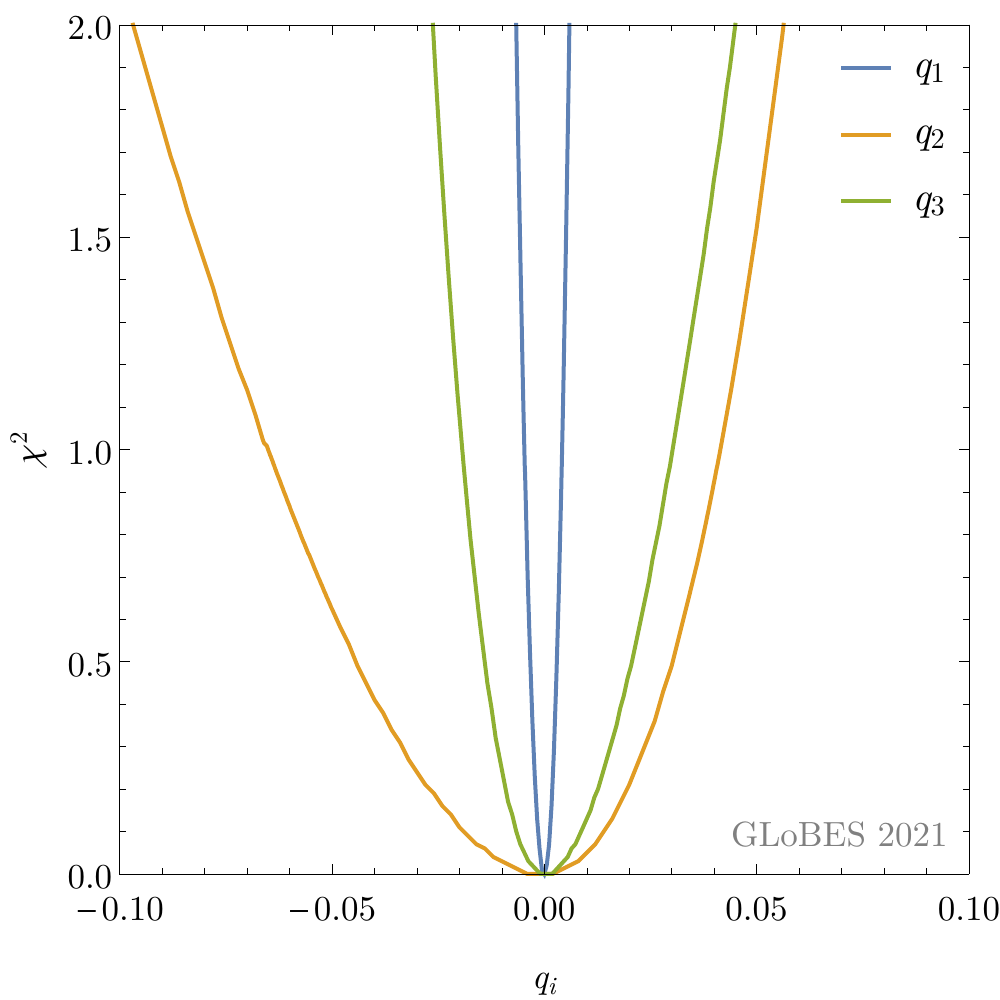}
	\caption{$\chi^2$ curves for $q_i$ as defined in Eq.~\ref{eq:qi} with all the systematic uncertainties listed in Ref.~\cite{Huber:2019frh}.}
	\label{fig:qchi2}
\end{figure}

\noindent
In order to compare the constraints for cases (1), (2), and (3), we define
\begin{align}
	& q_1\;\equiv\;\kappa\,\vev{\delta(E)}
	\,\text{,}\hspace{5mm}
	q_2\;\equiv\;\varepsilon
	\,\text{,} \qquad\text{and} \nonumber \\
	& q_3\;\equiv\; k\,\vev{1-P_\text{QM}(\nu_\alpha\to\nu_\beta)}
	\,\text{,}
	\label{eq:qi}
\end{align}
where $\vev{f(E)}$ is the average of $f(E)$ over the interval $1.8\,\text{MeV}\leq E\leq 8.0\,\text{MeV}$ using the oscillation parameters that minimized the $\chi^2$ at each point.  The results for analysis (a) including all systematic uncertainties is shown in Fig.~\ref{fig:qchi2}.

\section{Discussion}

In this letter, we point out the relevance of neutrino physics for addressing 
foundational questions in QM. 
In particular, we have examined the potential of the JUNO experiment
to probe for the triple-path interference among the three neutrino mass eigenstates and thereby test the Born rule.
The potential JUNO $1\sigma$ bound of $-0.017 < \kappa < 0.015$  
is competitive with those available from electromagnetic probes \cite{Sinha:2010,Park:2012}.
Moreover, the prospects of electromagnetic probes to improve their bounds
is limited due to the sensitivity of the interference pattern
on the change in boundary condition caused by the opening and closing of slits \cite{Yabuki:1986,PhysRevA.85.012101,PhysRevLett.113.120406,Sinha:2015,Rengaraj:2018},
whereas neutrino oscillations are independent of such considerations.

One drawback of our analysis is that we currently lack a theory which can
model departures from the Born rule, and predict how triple-path interferences
would depend on experimental parameters.
This is reflected in the arbitrary choices we must make to normalize
the triple-path interference $\varepsilon$ in our fits.
We are also assuming that triple-path interference will be introduced without any modification to the
existing double-path interferences, which may not be the case for a complete theory.
Without such a theory, we also cannot disentangle triple-path interference from other effects
such as neutrino non-standard interactions (NSI's), or the presence of sterile neutrinos.

A promising candidate theory that could potentially incorporate triple-path intereferences
within its framework is Nambu QM \cite{Minic:2002pd}.
This theory generalizes the time evolution of QM states 
in a way which is non-canonical yet unitary. 
In essence, Nambu QM generalizes the space in which the ``phase'' (in the sense of the ``phase'' of a complex number)
of a state is allowed to evolve, leading to non-canonical time evolution as well as
non-canonical double-path interference.
Indeed, we have recently discussed how the vanilla version of the theory
can be constrained using atmospheric neutrinos by looking at interference effects \cite{Minic:2020zjb}.
Given the larger freedom that the ``phase'' is allowed in Nambu QM,
we envision generalizations (most probably a non-associative one) in which the triple-path interference and
the departure from the Born rule could be precisely modeled.

We close this discussion by recalling that neutrinos can also probe 
the Leggett-Garg inequalities \cite{Formaggio:2016cuh,Fu:2017hky}, 
spontaneous collapse models of quantum measurement \cite{Donadi:2013uxa,Bahrami:2013hta},
non-standard time-evolution of quantum states \cite{Minic:2020zjb}, 
and, if M\"ossbauer neutrinos can be realized, 
the time-energy uncertainty relation \cite{Raghavan:2012sy}.
Further studies will indubitably lead to other ways to utilize neutrinos
for foundational QM studies.

\section{Acknowledgments}

We thank Chia Tze for helpful discussions. 
PH, DM, RP and TT are supported in part by the US Department of Energy (DE-SC0020262). 
DM is also supported by the Julian Schwinger Foundation.
TT is also supported in part by the US National Science Foundation (PHY-1413031).


\bibliographystyle{apsrev4-2}
\bibliography{Sorkin}

\end{document}